\documentclass[letter,traditabstract]{aa} 
\usepackage{graphicx}
\usepackage{natbib}
\usepackage{mwe,tikz}
\usepackage[percent]{overpic}
\usepackage{txfonts}
\def\igr{J17361}
\def\inte{{\em INTEGRAL}}
\def\xmm{{\em XMM-Newton}}

\def\chan{{\em Chandra}}

\def\swift{{\em Swift}}

\def \inte {{\em INTEGRAL}}
\def \xmm {{\em XMM--Newton}}

\def \hcm {\hbox {\ifmmode $ atom cm$^{-2}\else atom cm$^{-2}$\fi}}

\begin{document}
   \title{A $\sim$100~mHz QPO in the X-ray emission from IGR\,J17361-4441}

   \author{E. Bozzo
          \inst{1}
      \and A. Papitto
           \inst{2}
      \and C. Ferrigno 
           \inst{1}
      \and T. M. Belloni
            \inst{3}     
          }

   \institute{ISDC Data Centre for Astrophysics, Chemin d’Ecogia 16,
             CH-1290 Versoix, Switzerland; \email{enrico.bozzo@unige.ch}
              \and 
              Institut de Ci\`encies de l'Espai (IEEC-CSIC), Campus UAB, Fac. de Ci\`encies, Torre C5, parell, 2a planta, 08193 Barcelona, Spain
              \and  
              INAF, Osservatorio Astronomico di Brera, via E. Bianchi 46, I-23807 Merate (LC), Italy 
              }

   \date{}

  \abstract{IGR\,J17361-4441 was discovered by \inte\ undergoing its first detectable X-ray outburst in 2011 and initially classified as an accreting 
  X-ray binary in the globular cluster NGC\,6388. A reanalysis of the outburst data collected with \inte\ and \swift\ suggested  
  that the enhanced X-ray emission from IGR\,J17361-4441 could have been due to a rare tidal disruption event of a terrestrial-icy planet by a 
  white dwarf. In this letter we report on the analysis of \xmm\ data collected in 2011 during the outburst from IGR\,J17361-4441. 
  Our analysis revealed the presence of a 100~mHz quasi-periodic oscillation in the X-ray emission from the source and confirmed the presence 
  of a soft thermal component (kT$\sim$0.08~keV) in its spectrum. We discuss these findings in the context of the different possibilities  
  proposed to explain the nature of IGR\,J17361-4441. } 


   \maketitle

\section{Introduction}
\label{sec:intro}

IGR\,J17361-4441 (hereafter J17361) is a hard X-ray transient discovered by \inte\ \citep{winkler03} 
in the globular cluster NGC\,6388 \citep{gibaud11}. 
The first \inte\ detection of the source was reported on 2011 August 11, 
but the off-line analysis of \swift\,/BAT data revealed that the outburst might have 
started about 14~days before \citep[][hereafter S14]{delsanto14}.  
The total duration of the event was estimated to be $\sim$200~days \citep{bozzo12}. 
A \chan\ observation performed on 2011 August 29 revealed that the source was 
located outside the dynamic center of the globular cluster and thus ruled out the hypothesis that the dormant 
intermediate mass blackhole (IMBH) suspected to be hosted in NGC\,6388 could have experienced an episode of enhanced 
accretion \citep{lanzoni07,pooley11}. 

The duration of the X-ray outburst from \igr\ was reminiscent of what is typically 
observed from BH X-ray binaries. However, the stringent upper-limit obtained on the source radio emission  
seriously challenged this interpretation  
\citep{ferrigno11}. More controversial was the attempt to classify the source as a neutron star (NS) X-ray binary. 
On one hand, the localization of \igr\ within the globular cluster NGC 6388 ruled out the possibility of \igr\ being 
a high mass X-ray binary. On the other hand, NS low mass X-ray binaries (LMXBs) often display during 
outbursts pulsations down to millisecond periods and/or thermonuclear explosions.   
Neither of these phenomena were observed from \igr\ \citep{bozzo11}, thus leaving margins for 
alternative possibilities. 

An interesting interpretation of the nature of \igr\ was proposed by S14. These authors showed that the 
source X-ray luminosity decayed during the last $\sim$70~days of the outburst following a $t^{-5/3}$ profile that is typical 
of tidal disruption events (TDEs). Given the lack of any extended Galaxy in the HST observations performed in the direction of \igr,\ 
they proposed that the TDE was of Galactic origin and was caused by the disruption of a 
terrestrial-icy planet by a white dwarf (WD). As the latter has a typical accretion efficiency around 10$^{-4}$, the accreted mass 
during the entire outburst was estimated to be $\sim$2$\times$10$^{27}$~g. The authors also suggested 
that the soft thermal component with a temperature kT$\sim$0.08~keV and a radius $\sim$12600~km 
detected in the broad-band spectrum of the source (0.4-100~keV) was originated from the 
fall-back disk formed around the WD during the TDE. 

In this letter we report on the timing and spectral analysis of an \xmm\ observation 
that was performed about 42 days after the onset of the outburst from \igr.\    

\section{Data analysis}
\label{sec:data}
 
The \xmm\ \citep{jansen01} observation of \igr\ was carried out on 2011 September 23 for about 42.3~ks. The EPIC-pn was operated in timing mode, while the two 
MOS cameras were in full frame. We reduced these data by using the SAS version 13.5 and the latest calibrations files available on the 
\xmm\ repository\footnote{http://xmm2.esac.esa.int/external/ xmm\_sw\_cal/calib/index.shtml}. The observation was not affected by intervals of 
high flaring background, and thus we retained the entire exposure time available for the following analysis. For the EPIC-pn the source 
lightcurve and spectra were extracted in the energy range 0.6-12~keV\footnote{Data below 0.6~keV were discarded in order to avoid residual 
calibration uncertainties, see http://xmm2.esac.esa.int/ docs/documents/CAL-TN-0018.pdf} by selecting CCD columns comprised between 34 and 42. 
The background was extracted by using columns comprised between 3 and 14. We verified {\it a posteriori} that different reasonable 
choices of the source and background extraction regions did not affect significantly our final results. 
Two lightcurves of the source were extracted in the 0.6-1~keV and 1-12~keV energy bands. The lightcurves, background-subtracted 
and corrected for all relevant instrumental effects with the {\sc epiclccorr} task, were rebinned adaptively in order 
to achieve in each time bin a signal-to-noise ratio S/N$\geq$10 and calculate the corresponding hardness ratio  
\citep[HR; for the adaptive rebinning technique see][]{bozzo13}. No significant HR variations were measured during the observation; we 
thus extracted a single EPIC-pn spectrum by using all the exposure time available.      
\begin{figure}
  \includegraphics[width=5.5cm,angle=-90]{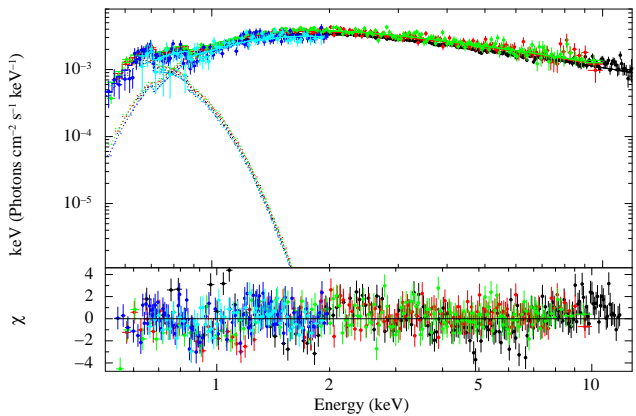}
  \caption{All spectra of \igr\ extracted during the \xmm\ observation. The EPIC-pn spectrum is represented in black, the MOS1 in red, the 
  MOS2 in green, and the combined RGS1 and RGS2 first (second) order spectra in blue (magenta). The best fit is shown, together with the residuals from 
  the fit.}    
  \label{fig:spectra}
\vspace{-0.5cm}
\end{figure}
EPIC-MOS1 and MOS2 data were extracted in the energy range 0.5-10~keV. 
Given the relatively high flux of the source, we found that the two MOS were significantly affected by 
pile-up\footnote{http://xmm.esac.esa.int/external/ xmm\_user\_support/documentation/ uhb/epicmode.html}. We used the tool {\sc epaplot} to 
estimate this effect and removed the pile-up by using for both cameras an annular extraction region centered around the source position with 
an external radius of 800 pixels ({i.e.} $40''$) and an inner radius of 280 pixels ({i.e.} $14''$). 
We also analyzed RGS data and extracted the source spectra by following standard 
procedures\footnote{http://xmm.esac.esa.int/sas/current/documentation/ threads/rgs\_thread.shtml}. 
Throughout this paper, uncertainties are given at 90\% c.l., if not stated otherwise. 

On the MOS images we noticed the presence of stray-light photons due 
to a bright source located outside the field-of-view. The stray-light features extended down to the boundary 
of the source point-spread-function. The above mentioned external radius of the MOS annular extraction region was chosen in order to avoid the 
largest contamination from the stray-light features. For the EPIC-pn data in timing mode, no stray-light correction is possible, as the camera 
provide limited spatial information in this operating mode. We note, however, that for a source as bright as \igr\  
the stray-light photons are usually considered to cause a negligible contamination in the EPIC and RGS data. 
The collecting area of these photons is indeed only a few cm$^2$ \citep{stockman98}, corresponding to about 0.2\% of the on-axis collecting 
area\footnote{See http://xmm.esac.esa.int/external/xmm\_user\_support/ documentation/uhb\_2.1/node23.html}.   
In order to verify that no significant contamination of the data took place, we performed both simultaneous fits with all EPIC and RGS 
spectra and with the MOS spectra alone, as the latter could be reasonably well corrected for the stray-light issue.   
The EPIC spectra were rebinned to have at least 25 photons per bin and prevent an oversampling of the energy resolution 
of the instruments by more than a factor of three. The RGS spectra were endowed with a relatively low statistics and no 
significant emission or absorption line could be revealed. We thus grouped these spectra in order to have at least 100 photons 
per bin and improve the S/N.  

We performed first a simultaneous fit of all \xmm\ spectra by using a simple absorbed ({\sc phabs} in {\sc Xspec}) 
power-law model (see Fig.~\ref{fig:spectra}). This fit gave an unacceptable result with $\chi^2_{\rm red}$/d.o.f.=4.1/633. We thus used the same spectral model proposed 
by S14 and added a {\sc diskBB} component with temperature $kT$$\simeq$0.08~keV and substituted the simple power-law with a {\sc cutoffpl}. 
This model provided a reasonably good description of the data ($\chi^2_{\rm red}$=1.44/631). Forcing a higher {\sc diskBB} 
temperature and lower normalization significantly worsened the fit ($\chi^2_{\rm red}$$\gtrsim$2). We fixed in the fit the 
value of the cut-off energy to that determined by S14 ($E_{\rm cut}$=41~keV) and note that leaving this parameter free to vary in the fit  
does not significantly affect the final results. We measured an average absorption column density of 
$N_{\rm H}$=(0.53$\pm$0.14)$\times$10$^{22}$~cm$^{-2}$, a power-law photon index of $\Gamma$=1.73$\pm$0.01, an inner disk temperature of 
0.082$\pm$0.002~keV, and a radius of (1320$^{+168}_{-145}$)$cos(\theta)$$^{-1/2}$~km, where $\theta$ is the inclination angle of the disk. 
We also included in the fit the normalization constants between different instruments. The normalization constant of the EPIC-pn 
was fixed at unity and we obtained $C_{MOS1}$=1.08$\pm$0.01, $C_{MOS2}$=1.13$\pm$0.01, $C_{RGS_1}$=0.92$\pm$0.02, and $C_{RGS_2}$=0.92$\pm$0.03 
for the normalization constants of the MOS1, MOS2, and the sum of the RGS first and second order spectra, respectively. 
Given the relatively limited energy band of the EPIC cameras, 
an equivalently good fit ($\chi^2_{\rm red}$=1.45/630) could be obtained by using a {\sc compTT} component instead of the cut-off power law  
(as suggested by S14). We did not measure significant changes in the properties of the {\sc diskBB} component within this model;  
the soft seed photons temperature of the {\sc compTT} component turned out to be compatible with the $kT$ reported above.    
The averaged unabsorbed (absorbed) 0.6-10~keV flux estimated from the spectral fit was 
5.0$\times$10$^{-11}$~erg~cm$^{2}$~s$^{-1}$ (3.0$\times$10$^{-11}$~erg~cm$^{2}$~s$^{-1}$). This corresponds to a luminosity of 
10$^{36}$~erg~s$^{-1}$ at a distance of 13.2~kpc (see S14 and references therein).  
We verified that the relatively large $\chi^2_{\rm red}$ of the fit was due to noisy spectral bins and no systematic trend 
appeared that could suggest the need for additional spectral components. 
Compatible results (to within larger uncertainties) 
are obtained by fitting only spectra from the two MOS cameras, proving that stray-light photons are not significantly contaminating  
the pn data.   

We also performed a timing analysis of all EPIC data. The source photons arrival times recorded by the 
three EPIC cameras were preliminary converted to the Solar system barycentre using the source
position determined by \chan\ \citep{pooley11}. As the EPIC-pn was
the only instrument giving a time resolution high enough
($3\times10^{-5}$~s) to search for periodic signals up to high
frequencies, we made use of the data from this camera to investigate the presence of coherent signals. 
No significant detection ($\gtrsim$3$\sigma$) was found in the frequency range 10$^{-3}$-150~Hz.  
In order to account for the possibility that \igr\ was part of a
binary system, we also searched for coherent signals by using shorter integration times for the power density spectra (PDS) of 3.96~ks. 
Even in this case, no significant detection was found and we estimated a 3$\sigma$ c.l. upper limit 
of 5.7\% on the amplitude of any coherent signal in the spanned frequency range. 

\begin{figure}
 \centering
   \begin{tikzpicture}[]
     \node at (1,1) {\includegraphics[width=6.0cm,angle=-90]{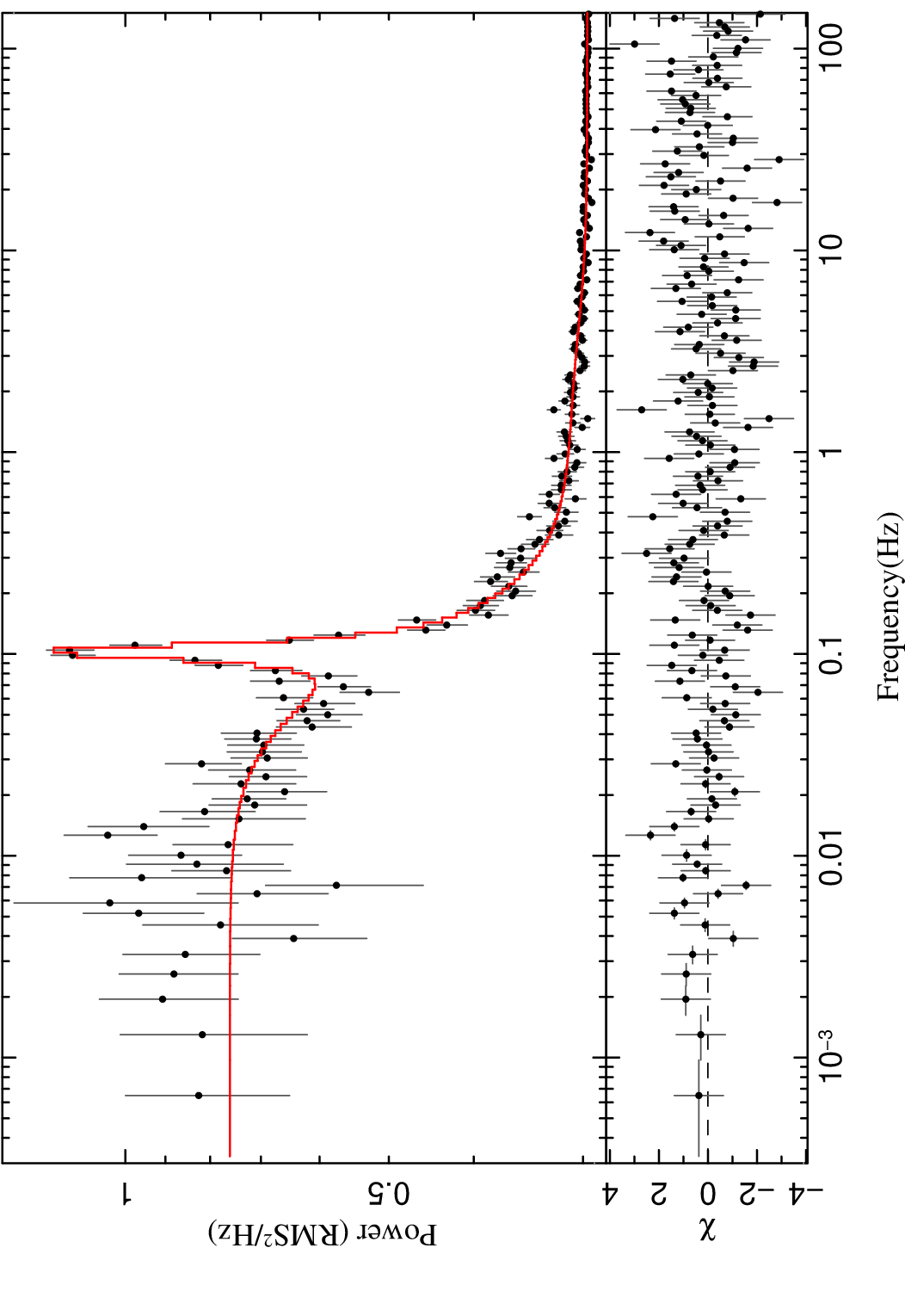}};
     \node at (3.05,2.35) {\includegraphics[scale=0.15,angle=-90]{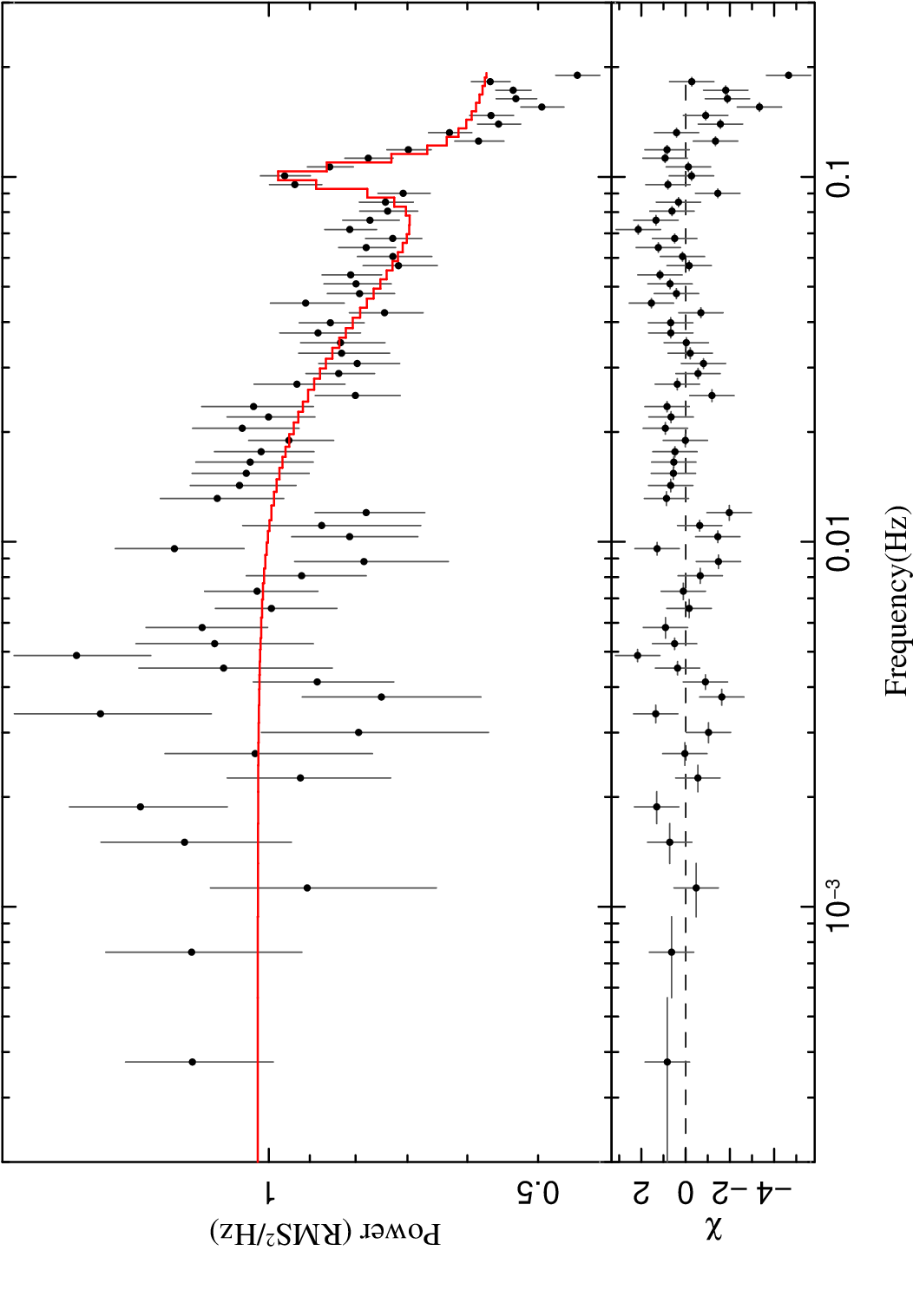}};
  \end{tikzpicture}
  \caption{Epic-pn PDS of \igr\ in the energy range 0.6-12~keV. The PDS has been rebinned 
  geometrically by using a factor of $1.05$. The red solid line represents the best-fit 
  model to the data (see Table~\ref{tab:timing}). The residuals from this fit are shown in the bottom panel. 
  Inset figure shows the PDS obtained from the two MOS, together with the  
  best fit and residuals. The QPO at $\sim$100~mHz is clearly visible also in this PDS.}    
  \label{fig:powersp}
  \vspace{-0.5cm}
\end{figure}

The EPIC-pn events (0.6-12~keV) were later binned every 2.952\,ms in order to compute a PDS in each stretch of $2^{19}$ bins  
(corresponding to roughly 1.55\,ks). All PDSs were then averaged to obtain the result shown in Fig.~\ref{fig:powersp}. 
The averaged PDS displayed a clear broad peak around 100~mHz. We tried to fit this PDS by using a model comprising a white noise component, 
a flat-top noise component and a quasi-periodic oscillation (QPO). As significant residuals were left from this fit, 
a second flat-top component was added to the model \citep[all noise components were modelled through Lorentian functions;][]{belloni02}. 
The best fit results are summarized in Table~\ref{tab:timing}.  

\begin{figure}
  \includegraphics[width=6.0cm,angle=-90]{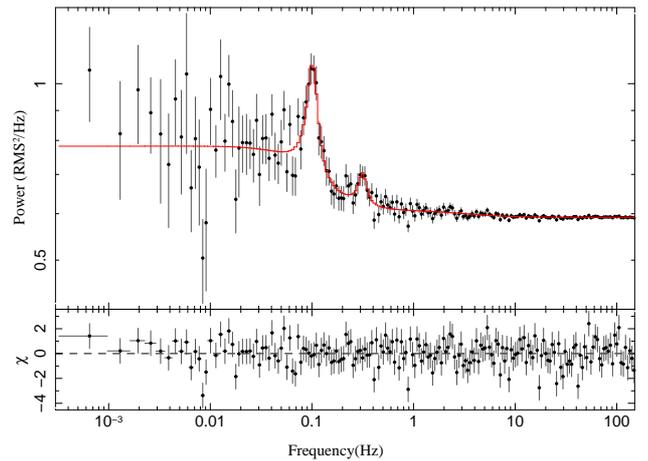}
  \caption{Same as Fig.~\ref{fig:powersp}, but the PDS here is extracted by using data in the 1.83-12.0~keV energy band.}    
  \label{fig:powersp2}
  \vspace{-0.1cm}
\end{figure}

\begin{figure}
  \includegraphics[width=2.6cm,angle=-90]{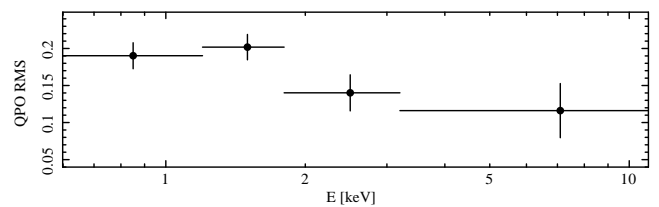}
  \caption{Energy dependence of the RMS amplitude of the $\sim$100~mHz QPO from \igr.\ }    
\label{fig:qporms}
  \vspace{-0.5cm}
\end{figure}

In order to search for possible energy dependence of source PDS, we also extracted the latter by dividing 
the EPIC-pn data in the two energy bands 0.6-1.83 and 1.83-12 keV (the energy bands were chosen in order to have roughly the 
same number of photons in each band). The QPO at $\sim$100~mHz was clearly visible in both PDSs. 
The PDS extracted in the softer energy band did not require the addition of 
the higher frequency flat-top noise component in the fit; the other PDS displayed the presence of a second QPO at $\sim$300\,Hz, consistent 
with being the third harmonic of the QPO at $\sim$100~mHz (see Fig.~\ref{fig:powersp2}). All the results obtained from the fits to the 
different PDSs are summarized in Table~\ref{tab:timing}. The energy dependence of the RMS for the QPO at $\sim$100~mHz is displayed 
in Fig.~\ref{fig:qporms}. The four energy bands were chosen to have a similar number of photons in each of them.  

We verified that the QPO at 100~mHz is also detectable when the event files from the two MOS cameras are merged together 
(see Fig.~\ref{fig:powersp}). 
A fit to the PDS extracted from the combined MOS data gave results compatible with those reported in Table~\ref{tab:timing} 
(in this case a model comprising a single flat-top noise component and a QPO at $\sim$100~mHz gave an acceptable fit, as the 
time resolution of the MOS in full frame is 2.6~s and the high frequency part of the PDS is limited to 0.2~Hz).  

\section{Discussion}
\label{sec:discussion}

The nature of the hard X-ray transient \igr\ remained so far elusive due to the lack of clear features in its X-ray emission 
that could help associating the source with one of the previously known class of objects displaying months-long X-ray outbursts 
reaching luminosities of $L_{\rm X}$$\gtrsim$10$^{37}$~erg~s$^{-1}$. 
The \xmm\ data of \igr\ that we analyzed in this letter provide further elements to investigate its nature. 

The \xmm\ observation was carried out about 42 days after the first \inte\ detection of the source, and took place during 
the $t^{-5/3}$ decay phase of the event as reported by S14. 
The combined fit of all EPIC and RGS spectra confirmed the emission model proposed by these authors, including 
a soft spectral component with a temperature of $kT$$\sim$0.08~keV. The radius of the thermal emission estimated by \xmm\ is significantly 
lower than that reported by S14; however, we note that their value was obtained from the spectral fit to the data extracted 
to within the first $\sim$8~days from the beginning of the outburst and no information is provided on the evolution of the {\sc diskBB} radius 
during the outburst. Furthermore, the absorption column density was fixed in their fit to 
0.8$\times$10$^{22}$~cm$^{-2}$, a value that is not fully consistent with that measured by \xmm.\ 
Even though such spectral model does not permit to fully establish the nature of \igr,\ we note that the parameters of the 
{\sc diskBB} component reported in Sect.~\ref{sec:data} would still be consistent with the suggestion that such emission is originated 
from the inner boundary of a WD fall-back disk, if the system is observed at a high inclination 
($\theta$$>$85~deg for a WD radius in the range 5-8$\times$10$^8$~cm).  

Our analysis also revealed the presence of a 100~mHz QPO in the X-ray emission from \igr.\ Similar features are commonly detected 
in several different classes of Galactic and extra-Galactic X-ray sources \citep[see, e.g.,]
[and references therein]{klis04}. LMXBs hosting accreting NSs so far displayed a wide variety of QPOs, spanning frequencies 
from a few up to $\sim$1300~Hz. These are associated to the motion of material within the NS accretion disk 
\citep{alpar85, lamb85, titarchuk99,lamb03}, and in some cases have been suggested to provide an efficient probe of 
general-relativistic effects \citep{stella98}. NS LMXBs hardly show QPOs at lower frequencies, the only exception being the 
case of mHz QPOs \citep{revnivtsev01}. However, these features appear around 7-9~mHz and are associated to thermonuclear explosions 
\citep{strohmayer03,altamirano08,linares12}, a phenomenon not observed from \igr.\  
Accreting WD in binary systems have long been known to display QPOs similar to those of NS LMXBs, and the frequency range spanned by 
these features includes the range of interest for the present analysis \citep{warner04}. 
QPOs from accreting WD are detected both in soft X-rays ($\ll$1~keV) and UV domain with periods either in the range 0.02-0.2~Hz 
or a factor of $\sim$10-20 larger\footnote{Longer-period DNOs were also observed in some cases \citep{warner04}}. The former features are usually called ``Dwarf Novae Oscillations'' (DNOs), as they are  
observed in the lightcurves of dwarf novae outbursts and have quality factors 10$^3$$<$Q$<$10$^7$; the latter have instead quality factors 
closer to those of QPOs in LMXBs (Q$\sim$5-20). It has been proposed that DNOs are the equivalent of the kHz QPOs in LMXBs, and thus such 
features are interpreted as being due to the motion of material close to the inner boundary of the accretion disk surrounding the WD 
\citep[see][and references therein]{warner02,wheatley03}. Indeed, as the Keplerian frequency of material orbiting in a disk 
is $\nu_{\rm k}$=1/2$\pi$($GM_{\rm CO}$/$R_{\rm CO}$$^3$)$^{1/2}$ ($M_{\rm CO}$ and $R_{\rm CO}$ are the compact object mass and radius), 
QPOs are expected at kilohertz for a NS system ($R_{\rm CO}$$\sim$10$^6$~cm) and at $\sim$100~mHz in the WD case  
($R_{\rm CO}$$\sim$5-8$\times$10$^8$~cm). 

Even though DNOs would have frequencies comparable with that of the QPO observed from \igr,\ the latter is detected 
up to much higher energies than those typical of DNOs. In particular, the QPO RMS fractional 
amplitude peaks at $\sim$1.2-1.8~keV and it is still significantly larger than zero up to 12~keV.  
As the thermal component with $kT$$\ll$1~keV detected from \igr\ was interpreted as being originated from the inner boundary 
of the accretion disk around the WD, it seems unlikely that such hard QPO can be formed in this region. 
As noticed by S14, the hard X-ray luminosity emitted during the event recorded from \igr\ would require 
anyway the presence of a hot corona in which hot electrons are able to up-scatter to higher energies the soft photons 
emitted from the disk through inverse-Compton.  
One could argue that, if the corona is located around the accretion flow and close to the disk inner boundary, 
it might provide the required environment where the soft QPO photons are up-scattered to higher energies. 
\begin{table}
\tiny
\caption{Parameters obtained from the fits to the EPIC-pn PDSs of \igr.\   
WN is the white noise component. The two flat-top components are zero-centered Lorentzians with full 
width half maximum (FWHM) $W_i$. $f_i$ is the QPO frequency and $Q_i$ its quality factor (where i=1,2). 
The latter is estimated by dividing the QPO central frequency by its FWHM.}
\begin{center}
\begin{tabular}{lccc}
 \hline
 \hline
 Range & 0.6-12 keV & 0.6-1.83 keV & 1.83--12 keV \\
 \hline
 WN & 2.055(1)       & 2.008(1) &  2.00(1) \\
 RMS$_{N1}$ (\%) &    15.4(7)   & 20.5(8)    & 12(2) \\
 $W_1$ (Hz)& 0.078(9)   & 0.060(7)   & 0.09(4) \\
 RMS$_{Q1}$ (\%) & 14.8(9) & 20(1)  & 13(2) \\
 $Q_1$ & 6(1)         & 6(1)    & 4(1) \\
 $f_1$ (Hz)& 0.102(1)    & 0.103(1) & 0.101(2) \\
 $RMS_{N2}$ (\%) & 20(2)     & -            & 20(3) \\
 $W_2$ (Hz) & 4(1) & -            & 4(2) \\
 RMS$_{Q2}$ (\%) & -          & -            & 9(2) \\
 $Q_2$ & -             & -            &  5(3) \\
 $f_2 (Hz) $ & -               & -           & 0.31(1) \\
 $\chi^2$/d.o.f.& 1.23/186  &  1.36/188        & 1.02/183\\ 
 \hline
\end{tabular}
\end{center}
\label{tab:timing}
\vspace{-0.7cm}
\end{table} 
It is worth noticing that the QPO revealed from \igr\ looks remarkably similar to other timing features 
observed from Ultra-Luminous X-ray sources \citep[ULXs;][]{strohmayer03,dewangan}. The nature of QPOs in these objects 
is still a matter of debate \citep{dheeraj}, but it has been proposed that they could be related to the motion of material 
close to the last stable orbit around a central BH. ULX QPOs have thus been used to weight the total mass contained 
in these systems and could provide support in favor of the IMBH model proposed to interpret the ULX nature. 
A similar interpretation holds for the $\sim$5~mHz QPO detected from the TDE around the supermassive BH Swift\,J164449.3+573451 
\citep{reis12}. The possibility of \igr\ being an extra-galactic source has been considered unlikely by S14  
due to its inclusion in the globular cluster NGC\,6388 and the lack of extended sources detected by HST in that   
direction.    

We thus conclude that, although it is difficult to completely rule out different possibilities for the nature of \igr,\ the 
results obtained from the analysis of the \xmm\ data are compatible with the TDE suggestion proposed by S14. 
Further observations of similar events with the next generation of large area X-ray instruments \citep[e.g., those on-board LOFT;][]{feroci14} 
will help understanding the nature of puzzling timing features, as that observed from J17361.

\section*{Acknowledgments}

TMB acknowledges support from PRIN INAF 2012-6 ``Accreting X-ray binaries: understanding physics through periodic and aperiodic variability". 
AP is supported by a Juan de la Cierva fellowship, and acknowledges grants AYA2012-39303, SGR2009- 811, and iLINK2011-0303.   

\bibliographystyle{aa}
\bibliography{J17361}

\end{document}